\documentclass[prd,amssymb,amsmath,amsfonts,nofootinbib,reprint,longbibliography,superscriptaddress, floatfix]{revtex4-2}
\usepackage{graphicx}
\usepackage{lmodern}
\usepackage{amsmath,amssymb}
\usepackage{mathrsfs}
\usepackage{amsfonts}
\usepackage[utf8]{inputenc}
\usepackage{url}
\usepackage[colorlinks]{hyperref}
\usepackage[dvipsnames,x11names,svgnames,rgb,table]{xcolor}
\usepackage{multirow}
\usepackage[normalem]{ulem}
\usepackage{float}
\usepackage{marvosym}
\usepackage{enumerate}
\usepackage{color,soul}
\usepackage{placeins}
\usepackage{bm}
\usepackage{color}
\usepackage{commath}
\allowdisplaybreaks
\usepackage
{multirow}
\usepackage{cleveref}
\usepackage{rotating} 
\usepackage{orcidlink}
\usepackage{todonotes}
\usepackage{makecell,tabularx,colortbl}
\usepackage{booktabs,cellspace}
\usepackage{mathtools}
\usepackage{aas_macros}

\definecolor{dodgerblue}{HTML}{1E90FF}
\definecolor{kgreen}{HTML}{009A4E}

\newcommand{\citeme}[1]{\textcolor{red}{[CITEME!]}}

\AtBeginDocument{
  \hypersetup{
    citecolor=kgreen,
    linkcolor=kgreen,   
    urlcolor=kgreen,
    filecolor=kgreen
    }
}

\newcommand{\bham}{\affiliation{School of Physics and Astronomy and Institute for Gravitational Wave Astronomy, University of Birmingham, Edgbaston, Birmingham, B15 2TT, United Kingdom}}

\begin{document}
\title{Prospects for characterizing Population III remnants with next-generation gravitational-wave observatories }
\author{N.~V.~Krishnendu \orcidlink{0000-0002-3483-7517}}
\email{k.naderivarium@bham.ac.uk} \bham
\author{Patricia Schmidt \orcidlink{0000-0003-1542-1791}}
\email{P.Schmidt@bham.ac.uk} \bham
\author{Geraint Pratten \orcidlink{0000-0003-4984-0775}}
\email{g.pratten@bham.ac.uk} \bham

\begin{abstract}
The most distant gravitational-wave (GW) detection by LIGO, Virgo and KAGRA so far is a binary black hole (BBH) merger at a redshift of $z \sim 1.1$, corresponding to a luminosity distance of $D_L \sim 8\, \rm Gpc$. The next-generation GW detectors, the Einstein Telescope (ET) and Cosmic Explorer (CE), will detect mergers beyond the peak of star formation at $z_{\rm peak}\sim 2$, enabling the direct detection of the remnants of the first stars in the early Universe. Realising this science potential requires accurate inference of the redshift, sky localization and intrinsic properties of the most distant mergers. 
In this work, using a fully Bayesian framework and an astrophysically motivated model, we study Population III remnants with an ET–CE detector network and quantify the measurement uncertainties in redshift, sky localisation, intrinsic masses and spins for BBHs at $z\geq15$. 
Considering an optimistic ($5\, \rm Hz$) and pessimistic ($10\, \rm Hz$) lower cutoff frequency for the detectors' sensitivity, we show that the $5\, \rm Hz$ configuration consistently improves the redshift inference for spin-precessing binaries. 
We also find that the source-frame component masses can be measured to within $\sim 12\%$ on average, and that the highest-redshift sources in the population can be reliably characterised. In contrast, we find only modest constraints on the BH spins. 
The improved low-frequency sensitivity also extends the redshift reach of the detector network, enabling events injected at $z_{\rm true}\simeq19.8$ to be confidently identified as originating beyond $z\simeq18.5$ at $90\%$ credibility, compared to a maximum lower-bound redshift of $z\simeq17.5$ for the $10\,\rm Hz$ configuration. Improved detector sensitivity below $10\, \rm Hz$ also reduces the sky-localization uncertainties which is essential for cosmological cross-correlation.

\end{abstract}
\date{\today}
\maketitle
%
%
\section{Introduction}
\label{intro}
The fourth observing run (O4) of the LIGO-Virgo-KAGRA (LVK) detector network~\cite{KAGRA:2013rdx, AdvancedLIGO2010, TheLIGOScientific:2014jea, TheVirgo:2014hva, TheVirgostatus,H1L1V1Dcc} successfully concluded on November 18, 2025, reporting approximately $250$ significant candidate events, bringing the total number of observed gravitational-wave (GW) signals to around $400$~\cite{GraceDB}.
The detections include binaries up to a redshift of $z\sim 1.1$ and a detector-frame total mass of $M^{\rm det}\sim 200 M_{\odot}$~\cite{GWTC-4-catalog-0,GWTC-4-catalog-1,GWTC-4-catalog-2,GWTC-4-catalog-3}.
While the LIGO, Virgo and KAGRA detectors are currently undergoing upgrades to improve their sensitivity for the next observation run, proposals for the next-generation (XG) of ground-based GW detectors, the Einstein Telescope (ET)~\cite{ET:2019dnz, ET:2025xjr, Punturo:2010zz} and Cosmic Explorer (CE)~\cite{Evans:2021gyd, Reitze:2019iox}, have now reached a mature state. 
These XG detectors will improve strain sensitivity by approximately an order of magnitude over current detectors and extend the sensitive frequency band, with particularly enhanced performance at both low $(\leq20 \rm Hz)$ and high frequencies. 
Improved low-frequency sensitivity is decisive for high-redshift binaries, as it increases the in-band duration of their signals as well as the accumulated signal-to-noise ratio (SNR), thereby improving detectability and parameter estimation accuracy. 

Recent studies have demonstrated that these improvements will enable forecasts of up to $\geq10^5$ binary detections per year extending to redshifts of order $z\sim100$~\cite{MPSAC, ET:2025xjr}.
Confidently detecting GWs from binary black holes (BBHs) at redshifts beyond the peak of star formation at $z_{\rm peak}\sim2$, offers a unique opportunity to explore black hole (BH) formation channels not traced by conventional stellar evolution~\cite{Ng:2020qpk}, in particular the formation of light seed BHs from the death of the first stars or direct collapse.

To date, the high-redshift BH population remains highly uncertain, while GW observations from an XG detector network promise to constrain their origin, evolution, and properties of the underlying astrophysical environment~\cite{Iacovelli:2022bbs,Plunkett:2025mjr}. 
In particular, such detections would provide direct constraints on their initial mass function, early-Universe star formation rate, and the resulting compact binary merger rate, thereby testing key predictions of early stellar structure and BH formation models. 
Access to this entirely new population would therefore not only help resolve existing tensions between theoretical predictions and observations, but also provide critical input for distinguishing between competing BH seed formation and growth scenarios that are inaccessible with low-redshift/late Universe observations alone~\cite{Fishbach_2018,Fishbach:2023pqs,GWTC-4-rates-pop}.

Complementary to GW observations, electromagnetic probes provide independent evidence for the existence of massive BHs in the early Universe. In particular, quasars powered by accreting BHs have recently been detected up to a redshift of $z \sim 10$~\cite{Natarajan:2023rxq}. 
Despite evidence for supermassive BHs $(\geq10^6M_{\odot})$ in the early Universe, their formation and rapid growth in low-metallicity environments remain unresolved~\cite{2023FanReview,2012Volonteri,Mangiagli:2023ize}.
Several formation scenarios have been suggested to explain the seeding and growth of supermassive BHs, namely, light, intermediate, and heavy seeds. 
These correspond respectively to the remnants of Population III (Pop III) stars, stellar-dynamical formation in dense clusters, and direct-collapse scenarios~\cite{2018Smith, Madau:2001sc, 2020Kroupa, Fryer:2000my, Begelman:2023cis}.
However, the absence of direct detections of BHs and BBH mergers at $z > 10$ continues to limit constraints on early-Universe compact object populations and formation channels~\cite{JWST2026_survey,JWST2026-2,Chira:2025jeu, 2023Keck, 2024ESO}.


The properties of BHs at these redshifts remain poorly constrained. 
The observed masses and spins encode the effects of stellar evolution and binary formation, while population-synthesis predictions are essential for connecting GW observations to the underlying stellar populations. 
Early studies of Pop~III binaries, for example, predicted detectable merger rates and characteristically massive systems~\cite{Kinugawa:2014zha,Kinugawa:2015nla,Hartwig:2016nde,Belczynski:2016ieo,Tanikawa:2020abs,Tanikawa:2021qqi,Ng:2022agi}, while subsequent work also considered dynamical formation in Pop~III star clusters~\cite{Liu:2020ufc}. 
More recently, Ref.~\cite{Costa:2023xsz} used semi-numerical models spanning different assumptions about the initial mass function and orbital properties, finding only mild differences between the mass distributions of Pop~III and  BBHs formed from Ppoulation~II stars.
Subsequent studies quantified how uncertainties in Pop III star-formation histories and binary orbital properties propagate into merger-rate predictions~\cite{Santoliquido:2023wzn}. 
These uncertainties can lead to variations of up to two orders of magnitude and shift the merger-rate peak to $z \sim 8–16$~\cite{Santoliquido:2023wzn}. 
Using parametric and non-parametric astrophysical population models and full Bayesian inference, Ref.~\cite{Plunkett:2025mjr} estimated the population-level features of Pop I/II and Pop III BBHs, finding that XG detectors are capable of constraining the logarithm of the star formation rate density to within $\sim25\%$ over redshifts $10 -20$.

These studies highlight that the astrophysical uncertainties in early star formation and binary evolution models dominate the predicted high-redshift merger population, underlining the need for observational constraints from GWs. 
Reference~\cite{Ng:2020qpk} showed that through large numbers of high-SNR BBH observations from XG GW detectors, it is possible to reconstruct the merger-rate evolution and simultaneously infer multiple formation channels, including a potential high-redshift ($z \gtrsim 5$) population as a signature of unconventional formation.

However, the ability to detect a high-redshift binary population does not automatically guarantee that the physical properties of individual sources can be reliably inferred, as parameter degeneracies can still lead to large uncertainties in progenitor properties~\cite{Ng:2021sqn,Ng:2022vbz,Vitale_2021}. 
To quantify this issue, Ref.~\cite{Mancarella:2023ehn} introduced an inference-based framework that evaluates whether individual sources can be meaningfully characterised in addition to being detected, by assessing the expected accuracy of parameter recovery using a mock population and the Fisher matrix formalism.

In this work, we perform full Bayesian inference on an astrophysically motivated Pop III remnant population to assess the extent to which such high-redshift BBH sources can be reliably characterised beyond detectability, by an XG detector network. 
A key requirement for the exploration of high-redshift BHs is access to GW frequencies below $20$ Hz. Enhanced design choices such as longer arms, improved Newtonian noise mitigation and seismic isolation as well as quantum-noise suppression and new coatings are key to unlocking access to the crucial $3-20$ Hz regime with XG observatories. Achieving the projected sensitivity below $10$ Hz will be challenging but is critical for constraining the high-redshift Universe. 
To quantify the importance of the enhanced low-frequency sensitivity, we show direct comparisons considering $10\rm Hz$ and $5 \rm Hz$ for the lower cut-off frequency.


Focusing on binaries at redshift $z\geq 15$ and employing a state-of-the-art phenomenological waveform model for binaries with spin-induced precession effects and higher-order modes, we determine the capabilities of XG detectors in placing meaningful constraints on the source redshift, mass, and spin distributions along with discussing the sky localization uncertainties, a critical information for follow-up and cosmological cross-correlation. In Sec.~\ref{sec:methodlogy} we describe the initial population properties of the Pop III star remnants, the considered XG detector configurations, and the Bayesian inference implementation. We discuss our key results in Sec.~\ref{sec:results}, and we conclude in Sec.~\ref{sec:conclusions}.

\begin{figure*}[t]
    \includegraphics[width=0.48\textwidth]{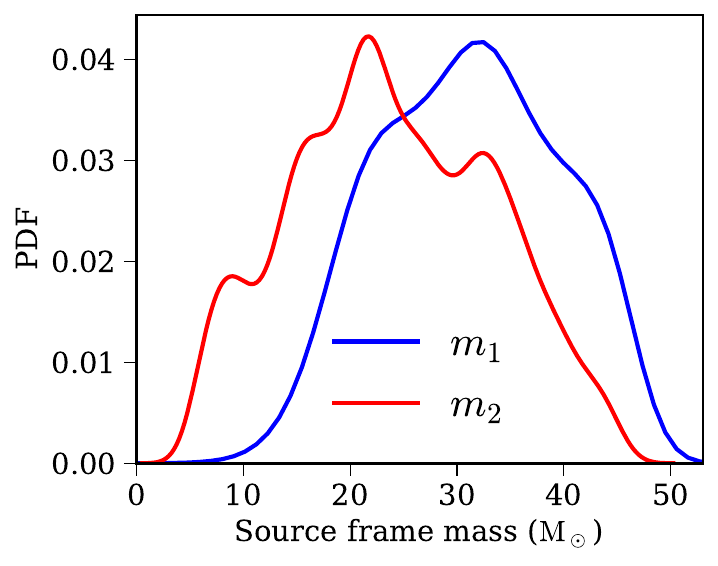}
    \includegraphics[width=0.48\textwidth]{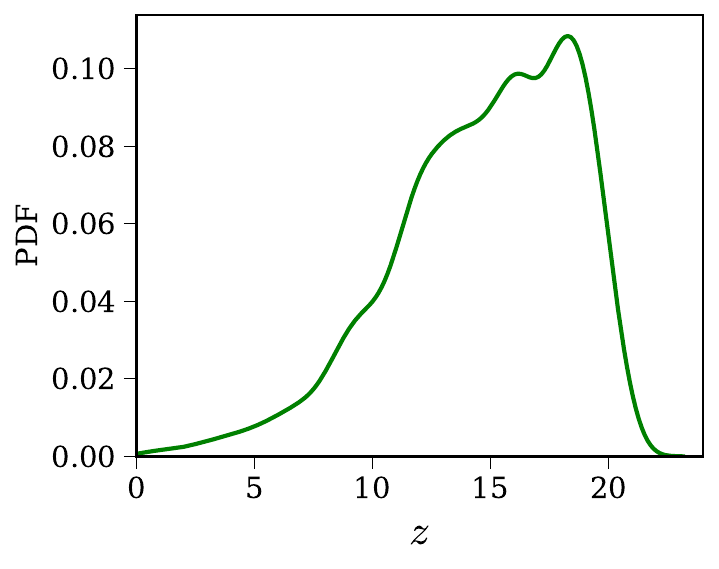}
    \caption{Probability distribution function of the source-frame component masses $m_1$ (blue) and $m_2$ (red), and the redshift distribution of the Pop III remnants following the LOG IMF and SW20 SFRD model in~\cite{Santoliquido:2023wzn, Santoliquido:2024oqs}. The initial population is generated by drawing independent random samples for the primary mass, secondary mass, and redshift from their respective distributions. Component spins are sampled assuming uniform magnitude and isotropic orientations. The source location, orientation, and polarization angles are drawn from isotropic and uniform distributions.
    }
    \label{fig:initial_pop_source_frame_massses_redshift}
\end{figure*}
\begin{figure}
    \includegraphics[width=0.48\textwidth]{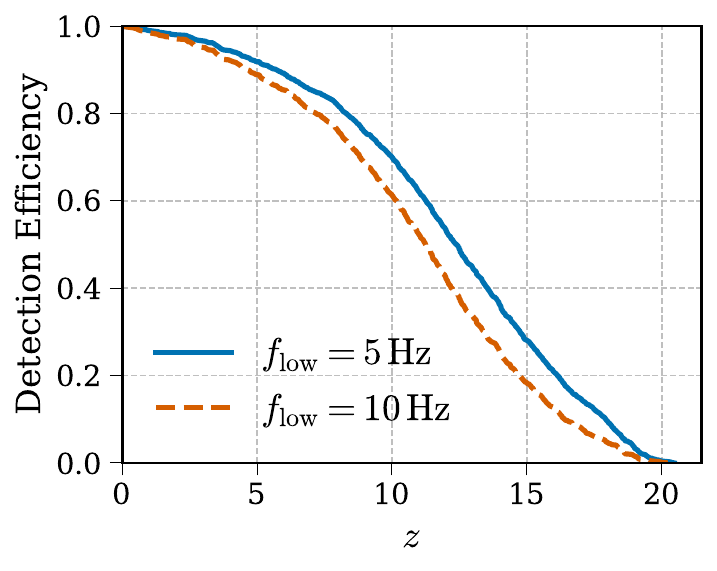}
    \caption{Detection efficiency of the XG detector network as a function of redshift. The efficiency is computed as the ratio of detected to injected Pop III BBH mergers in each redshift bin. A source is considered detected when the SNR in each individual detector exceeds $8$ and the network SNR exceeds $30$. Results are shown for analyses performed with low-frequency cutoffs of $f_{\rm low}=5\, \mathrm{Hz}$ and $f_{\rm low}=10\, \mathrm{Hz}$. Lowering the low-frequency cutoff increases the fraction of detectable sources 47.1\%.}
    \label{fig:det_efficiency}
\end{figure}
\section{Methodology}
\label{sec:methodlogy}
%
Although XG detectors can detect BBHs out to high redshifts, robust parameter inference is limited to a much smaller range than the nominal detection horizon. Using Fisher-matrix error estimates and a mock BBH population, Ref.~\cite{Mancarella:2023ehn} introduced two diagnostics to quantify the distinction between detectability and parameter inference of BBHs: (i) the inference horizon and (ii) the $z$--$z$ consistency plot; the former defines the maximum redshift at which a specified fraction of sources can be characterised to a given accuracy, while the latter compares inferred and true redshifts to quantify the bias and scatter in the recovered values. The analysis of Ref.~\cite{Mancarella:2023ehn} is limited by two assumptions: the use of a mock BBH population and the Fisher-matrix approximation for parameter estimation. A more realistic assessment of XG inference capabilities requires astrophysically motivated population models together with a fully Bayesian treatment, which naturally captures parameter degeneracies and the complete likelihood structure.

\begin{figure}
    \includegraphics[width= 3. in]{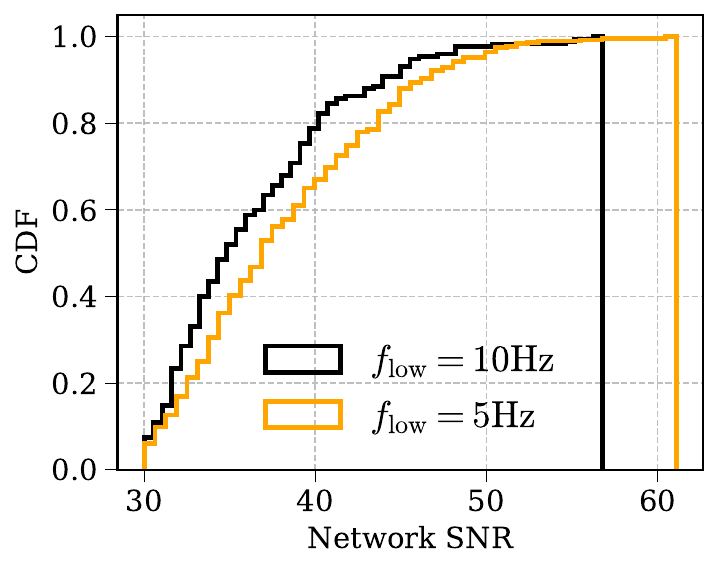}
    \caption{Network SNR distributions for a three-detector network comprised of two CE detectors (CE20 and CE40) and a triangular ET detector. The sources are selected based on a detection criterion that requires an individual-detector SNR exceeding $8$ and a network SNR exceeding $30$. These thresholds are applied to all early-Universe binaries located at redshifts $z \geq 15$, which pass both the individual-detector and network SNR criteria.
    }
    \label{fig:snr_dist_flow_comp}
\end{figure}

\subsection{Astrophysical Model of Pop III Remnants}
\label{subsec:popIII}
We model the Pop III star remnant population following Santoliquido et al.~\cite{Santoliquido:2023wzn, Santoliquido:2024oqs}, using semi-analytic prescriptions for non-spinning BBHs formed via Pop III stellar collapse. The framework combines Pop III stellar evolution tracks from the PARSEC code~\cite{PARSEC}, binary population synthesis with the SEVN code~\cite{Iorio:2022sgz}, and redshift-dependent merger-rate densities computed with CosmoRate~\cite{Santoliquido:2020axb}. Reference~\cite{Santoliquido:2023wzn} explores a range of initial mass functions (IMFs) and Pop III star formation rate density (SFRD) models, spanning different assumptions on stellar mass distributions, binary orbital properties, and cosmic star formation histories. Each SFRD model includes self-consistent distributions of binary parameters, including mass ratio, orbital period, and eccentricity, enabling a systematic assessment of the resulting BBH population and its detectability with future GW detectors.
In this work, we adopt the LOG IMF and the SW20 SFRD as the fiducial model. The LOG IMF provides a phenomenological stellar mass distribution that enhances the fraction of massive stars, while the SW20 SFRD model describes the cosmic star-formation rate density as a function of redshift, calibrated to observational constraints.

Figure~\ref{fig:initial_pop_source_frame_massses_redshift} shows the source-frame mass distributions of the primary and secondary BHs in the binary, with $m_1 \geq m_2$, and the initial redshift distribution (right). The source-frame mass distribution is dominated by binaries with source frame component masses in the range $20$–$40\,M_{\odot}$, and the redshift distribution peaks at $z \sim 18$~\cite{Santoliquido:2023wzn}.
We note that the maximum redshift is model-dependent; in this case, the distribution spans $z\sim 0-22$. For the subsequent analysis, we apply a redshift-based selection criterion and restrict the population to binaries with $z \geq 15$, targeting the early Universe population where star formation is expected to be dominated by metal-poor Pop III stars.

While the component mass and redshift distributions are from~\cite{Santoliquido:2023wzn}, the spin magnitudes are independently and identically drawn from a uniform distribution in the range $[0,1]$, and spin orientations are assumed to be isotropically distributed for each binary in the initial population. We do not impose a mass pairing function but pair random draws requiring $m_1 \geq m_2$.
Sky location, orbital orientation, and polarization angles are sampled uniformly over their full ranges. 
We randomly draw $10^3$ binaries from these distributions, which ensures a sufficiently large detected sub-sample for robust population-level binary parameter inference while keeping computational costs tractable. Given the poorly constrained merger-rate estimates of these high-redshift binaries, this choice is made to ensure a sufficient number of detected sources for constraining the properties of detected population.
Further details on the selection criteria, detector configuration, and resulting observable population are provided in Sec.~\ref{sec:configs}.

\subsection{Detector Network and Observable  BBH Population}
\label{sec:configs}
\begin{table}[t]
\centering
\renewcommand{\arraystretch}{1.4}
\begin{tabularx}{\linewidth}{l X}
\hline
\multicolumn{2}{c}{{Initial Population \& detector configuration}} \\
\hline
XG network & ET + CE \\
ET configuration & Triangular configuration with $10\,\mathrm{km}$ arms, located in Sardinia, Italy \\
CE configuration & Two detectors with $20\,\mathrm{km}$ (Texas) and $40\,\mathrm{km}$ (Washington) arm lengths \\
Initial population & $10^{3}$ binaries \\
Redshift selection & $z \geq 15$ \\
SNR thresholds & Single-detector: $8$; Network: $30$ \\
Low-frequency cutoff & $f_{\mathrm{low}} = 5\,\mathrm{Hz}$ and $10\,\mathrm{Hz}$ \\
Detected binaries ($5\,\mathrm{Hz}$) & $395$ \\
Detected binaries ($10\,\mathrm{Hz}$) & $175$ \\
Noise PSD & Taken from~\cite{MPSAC} \\
\hline
\end{tabularx}
\caption{Summary of the XG detector network configuration and initial population of Pop III remnants.}
\label{tab:xg_summary}
\end{table}

We consider an XG detector network consisting of ET and CE. The ET is assumed to be in a triangular configuration located in Sardinia, Italy, with $10\mathrm{km}$ arm length  in the xylophone tuning~\cite{ET:2019dnz, MPSAC}. The CE network comprises of two detectors with arm lengths of $20\mathrm{km}$ (CE20) and $40\mathrm{km}$ (CE40), placed at fiducial locations in Texas and Washington, respectively. To assess the performance of the XG network in constraining Pop III star remnant properties, we consider two low-frequency cutoffs: $f_{\mathrm{low}} = 5\, \mathrm{Hz}$ and $f_{\mathrm{low}} = 10\, \mathrm{Hz}$, which represent optimistic and conservative assumptions for the low-frequency sensitivity.

From the initial population of $10^{3}$ randomly drawn binaries, we select systems with $z \geq 15$ and require the single-detector SNR to be $\geq 8$ and the network SNR to be $\geq 30$ for detection.
This ensures that the subset of binaries adopted in this study are sufficiently informative for the question at hand. 
Out of the initial sample, $395$ binaries satisfy these criteria for $f_{\mathrm{low}} = 5\, \mathrm{Hz}$, and $175$ binaries are detected for $f_{\mathrm{low}} = 10\, \mathrm{Hz}$. 
Because both low-frequency cut-offs are applied to the same injection set using identical selection criteria, access to the $5$--$10$ Hz frequency band increases the number of sources passing our high-SNR threshold by a factor $\sim 2.3$. 
%
The noise spectral densities for both detectors are taken from Ref.~\cite{MPSAC}. A summary of the detector configurations and detected binary population is given in Tab.~\ref{tab:xg_summary}.

Figure~\ref{fig:det_efficiency} shows the detection efficiency, defined as the fraction of injected sources recovered in each redshift bin, $\epsilon(z)=N_{\rm det}(z)/N_{\rm inj}(z)$, where $N_{\rm det}$ and $N_{\rm inj}$ denote the numbers of detected and injected sources,
as a function of redshift. 
The blue solid and orange dashed curves correspond to analyses performed with low-frequency cutoffs of $f_{\rm low}=5\,\mathrm{Hz}$ and $f_{\rm low}=10\,\mathrm{Hz}$, respectively. Lowering the low-frequency cutoff increases the fraction of detectable systems by 47.1\%.

Figure~\ref{fig:snr_dist_flow_comp} shows the resulting SNR distributions of the detected binaries. 
For both low-frequency cut-offs, the network SNRs span a broad range, extending from approximately 30 to 60. Roughly 40\% of the sources have network SNRs below $\sim35$, while about 80\%--90\% of the sources lie below SNR~$\sim45$. The overall SNR distribution for the $5\,\mathrm{Hz}$ and $10\,\mathrm{Hz}$ configurations are very similar, with only a modest shift toward higher SNRs for the $5\,\mathrm{Hz}$ case. The most noticeable difference appears in the high-SNR tail, where the $5\,\mathrm{Hz}$ configuration contains the loudest sources, reaching a maximum network SNR of $\sim61$ compared to $\sim57$ for the $10\,\mathrm{Hz}$ configuration. These results confirm that extending the analysis bandwidth down to $5\,\mathrm{Hz}$ systematically increases the network SNR across the population.
%
\begin{figure*}
    \includegraphics[width=0.48\textwidth]{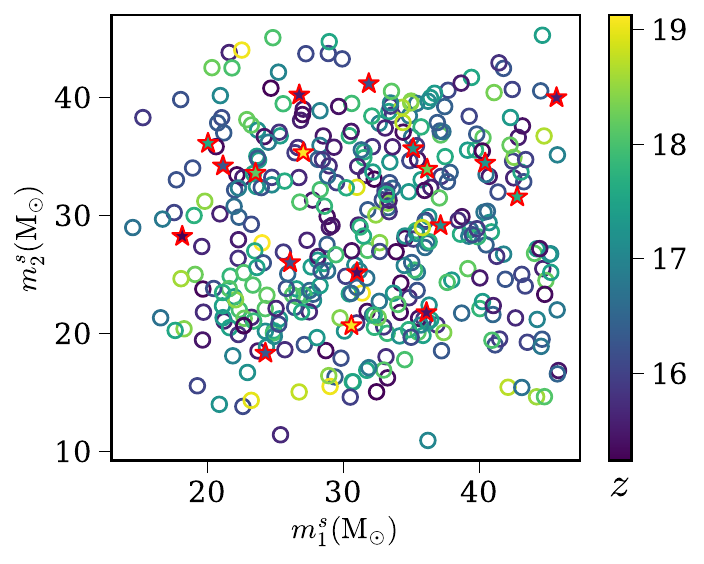}
    \includegraphics[width=0.48\textwidth]{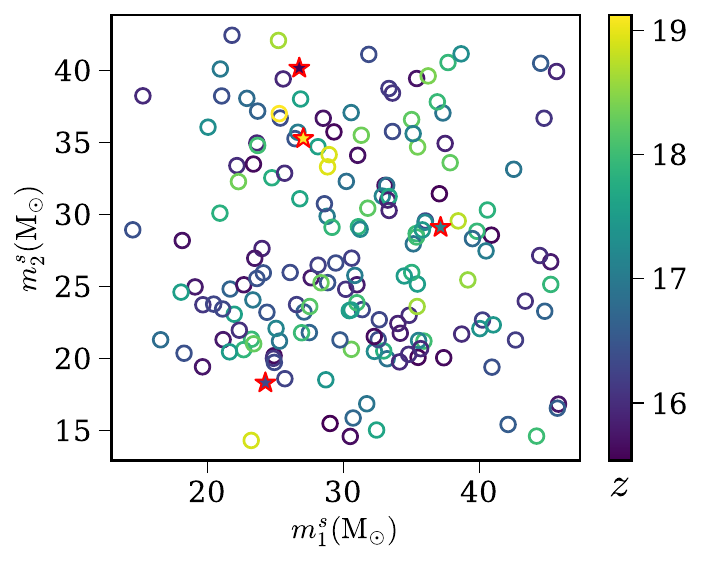}
    \caption{Left: The source-frame total mass distribution of binaries satisfying the detection threshold and their corresponding redshift for a low-frequency cutoff of $f_{\mathrm{low}} = 5\mathrm{Hz}$. For this configuration, 18 sources yield network SNRs above 50 in the XG detector network. Right: The same quantities are shown for a low-frequency cutoff of $f_{\mathrm{low}} = 10\mathrm{Hz}$. In this case, 4 events produce network SNRs above 50. In both panels sources with a network SNR $> 50$ are marked by an asterisk. 
    }
    \label{fig:m1_m2_scatter_5Hz}
\end{figure*}

The source-frame component masses of the detected BBHs are shown in Fig.~\ref{fig:m1_m2_scatter_5Hz}, with the redshift indicated by the color bar. Among the 395 and 175 sources in the two cases, binaries with network $\mathrm{SNR}>50$ are marked with red asterisks. There are 18 such events for $f_{\rm low}=5\,\mathrm{Hz}$ and only 4 for $f_{\rm low}=10\,\mathrm{Hz}$.
For the detected sources, we perform Bayesian inference with the \texttt{IMRPhenomXPHM} waveform model to determine how accurately we can estimate the key properties of the high-redshift binaries.

\begin{figure}
    \includegraphics[width=0.48\textwidth]{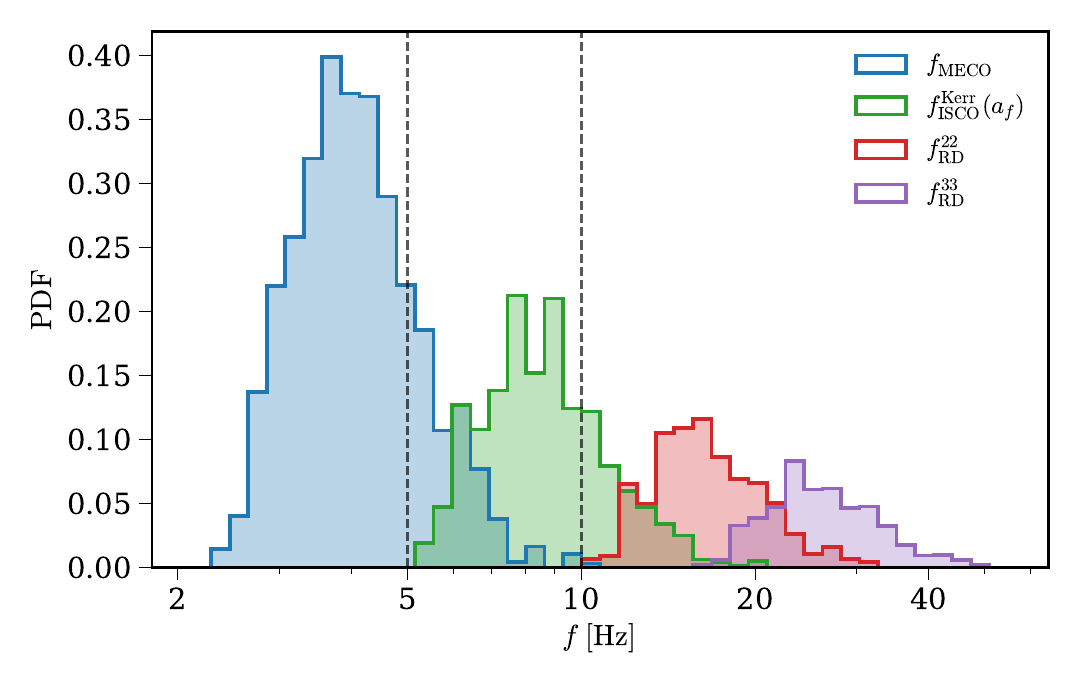}
    \caption{Distributions of the characteristic frequencies of the detected Pop~III remnant population: the MECO frequency $f_{\rm MECO}$ marking the end of inspiral~\cite{Pratten:2020ceb}, the Kerr ISCO frequency of the remnant $f_{\rm ISCO}$~\cite{Ori:2000zn}, and the $(2,2)$ and $(3,3)$ ringdown frequencies~\cite{Garcia-Quiros:2020qpx}. 
    Vertical lines indicate the two low-frequency cutoffs considered in this analysis.
    For most sources the inspiral terminates below $5\,\mathrm{Hz}$, while the ringdown lies above $10\,\mathrm{Hz}$ for the entire sample.}
    \label{fig:char_freqs}
\end{figure}

The role of the low-frequency cutoff is qualitatively different for this
population than for stellar-mass binaries at low-to-moderate redshifts. 
The choice of the low-frequency cutoff effectively determines which stages of the BH coalescence enter the sensitivity band of the detectors. 
The selected sources have detector-frame total masses of $M^{\rm det}\simeq740$--$1480\,M_{\odot}$, placing the end of the observed inspiral at $f_{\rm MECO}\simeq2.9$--$6.7\,\mathrm{Hz}$ and the $(2,2)$ ringdown at $f_{\rm RD}^{22}\simeq12$--$27\,\mathrm{Hz}$~\cite{Pratten:2020ceb}, as shown in Fig.~\ref{fig:char_freqs}.
A cutoff of $10\,\mathrm{Hz}$ is broadly insensitive to the inspiral regime for this population of high-redshift binaries, restricting the observable signal to the late merger and ringdown. 
Lowering the cutoff frequency to $5\,\mathrm{Hz}$ recovers substantially more of the merger and, for the $\sim30\%$ of sources with $f_{\rm MECO}>5\,\mathrm{Hz}$, portions of the late inspiral. 
However, these statements are only based on the dominant $(2,2)$ harmonic. 
The higher-order multipoles have frequencies that scale as $f_{\ell m} \sim (m/2) f_{22}$~\cite{Garcia-Quiros:2020qpx}, shifting the inspiral and merger-ringdown content to higher frequencies. 
As a result, the higher-order multipoles can start to generate a significant fraction of the accessible signal for these binaries.

\subsection{Bayesian Inference}
\label{sec:pe}
A quasi-circular BBH is characterised by the individual masses $m_i$, spin vectors $\bm {S}_i$, luminosity distance $D_{\rm L}$, angles specifying the location in the sky, $(\alpha, \delta)$, the orientation of the source, $\theta_{\rm JN}$ and the  polarization angle $\psi$, denoted by the set $\bm \theta$, 
\begin{equation}
    \bm \theta \equiv \{m_i, \bm{S}_i, D_{L}, \alpha, \delta, \theta_{\rm JN}, \psi, t_c, \phi_c\},
\end{equation}
where $i=1,2$ labels the primary and secondary, and the two nuisance parameters $t_c$ and $\phi_c$ are the reference time and phase, respectively.
We estimate the posterior probability distributions of the binary parameters using Bayesian inference~{~\cite{Veitch:2009hd,Veitch:2014wba, Skilling:2006gxv}}. 
This multi-dimensional posterior probability distribution function is proportional to the likelihood $\mathcal{L}$ and the prior $\pi$, given by
\begin{equation}
p(  \bm \theta | d ) \propto  \mathcal{L}(d | \bm \theta) \, \pi( \bm \theta),
\label{eq:post}
\end{equation}
where $d$ denotes the observed GW data. 
The one-dimensional distribution for any parameter $\theta \in \bm \theta$ is obtained by marginalizing over the unwanted parameters, 
\begin{equation}
    p( \theta | d ) = \int p(\hat{\bm \theta}, \theta|d) d\hat{\bm\theta}.
\label{eq:post-marg}
\end{equation}
From the marginalized posteriors we define credible intervals, which provide a quantitative measure of the uncertainty of the inferred parameters.
For a given parameter $\theta$, the symmetric $X\%$ credible interval (CI) is constructed by identifying the interval $[\theta_{\rm low}, \theta_{\rm high}]$ such that
\begin{equation}
\int_{\theta_{\rm low}}^{\theta_{\rm high}} p(\theta|d)d\theta = X/100,
\label{eq:CI}
\end{equation}
with equal probability contained in the lower and upper tails of the distribution.
%
Relative uncertainties on $\theta$ are computed with respect to the median $\theta_{\rm med}$ as
\begin{equation}
    \frac{\Delta \theta}{\theta} \equiv \frac{\theta_{\rm high} - \theta_{\rm low}}{2\,\theta_{\rm med}}.
\label{eq:relerror}
\end{equation}
To assess systematic offsets between the posterior and the true parameter values, we compute the posterior quantile diagnostic $Q$, defined as
\begin{equation}
Q(\theta_{\rm true}) \equiv \frac{1}{2} - \int_{\theta_{\min}}^{\theta_{\rm true}} p(\theta \mid d)\, \mathrm{d}\theta,
\label{eq:Qdef}
\end{equation}
where $p(\theta \mid d)$ is the posterior distribution of the parameter $\theta$ given the data $d$, and $\theta_{\min}$ is the lower bound of the prior support.
In practice, $Q$ is estimated from posterior samples as
\begin{equation}
Q(\theta_{\rm true}) = \frac{1}{2} - \hat{F}(\theta_{\rm true}),
\end{equation}
where $\hat{F}$ is the empirical cumulative distribution function of the samples, evaluated at the true value.
By construction, $Q \in [-1/2, 1/2]$.
A value of $Q = 0$ indicates that the true value coincides with the posterior median, while $Q > 0$ ($Q < 0$) indicates systematic overestimation (underestimation) of the parameter~\cite{Pratten:2020igi}. 
Likewise, the magnitude $|Q|$ describes the posterior probability mass between the posterior median and the injected value. 
Since $Q$ depends jointly on the location, width, and shape of the posterior, it is a useful diagnostic of biases in estimates of the source parameters, rather than of absolute error or precision.

Sampling is performed with the nested sampling algorithm \texttt{Dynesty}~\cite{DYNESTY}, implemented in \texttt{Bilby}~\cite{Ashton:2018jfp}.
We use a uniform prior in the range $[100, 1200]M_{\odot}$ for the detector-frame component masses, with slight extensions in a few cases to avoid boundary effects. 
However, a prior that is uniform in the detector-frame masses is
informative in the source-frame quantities. 
Because $m^{\rm src} = m^{\rm det}/(1+z)$, it induces a non-uniform prior on $(m^{\rm src}, z)$ and the resulting source-frame mass posteriors inherit a redshift-dependent weighting.
We adopt a uniform prior $\in [0,1]$ for the spin magnitudes, while spin orientations are assumed to be isotropically distributed. 
Luminosity distance is chosen as the sampling parameter and the corresponding redshift distribution is calculated assuming the Planck15 cosmological parameters~\cite{GWTC-4.0-catalogue-1,Ade2016,LIGOScientific:2025jau}. The luminosity distance prior is given by a power-law with index $\alpha =2$, $\pi(d_{L})\sim d_{L}^2$, with a range of $\mathrm{[1, 1000]\,Gpc}$. To account for physically motivated cosmological effects in the redshift estimation, we resample the posterior samples using a prior that corresponds to a uniform merger rate per comoving volume in the source frame, assuming a flat $\Lambda\mathrm{CDM}$ cosmology~\cite{GWTC-2-catalogue, Planck:2015fie}. 
See Appendix~\ref{ap:resam} for more details of the resampling procedure. 
Unless stated otherwise, all redshift results quoted in this work are obtained from the reweighted posteriors.

We assume the \texttt{IMRPhenomXPHM} waveform model for both injection and recovery to avoid any biases due to waveform systematics. All injections are performed in zero noise assuming the PSDs listed in Tab.~\ref{tab:xg_summary}. The sampling rate is 4096 Hz and the likelihood integration is performed between $f_{\rm low} = 5$ Hz or $10$ Hz and the Nyquist frequency. 

\begin{figure*}
    \includegraphics[width=0.48\textwidth]{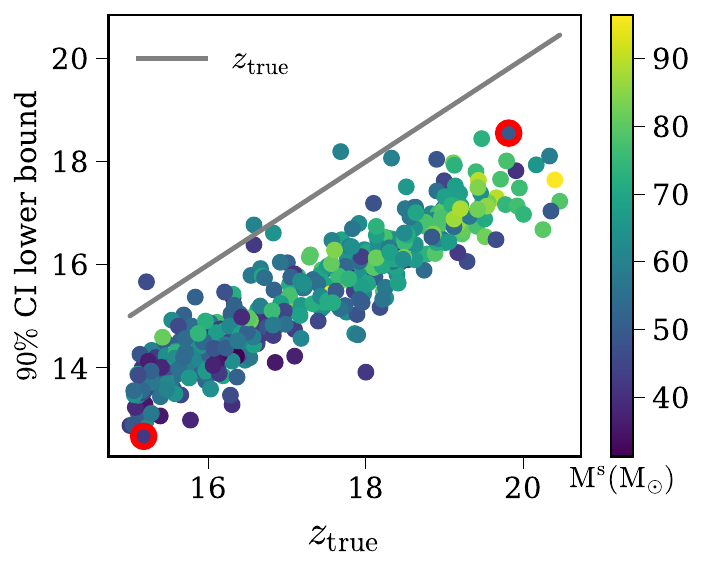}
    \includegraphics[width=0.48\textwidth]{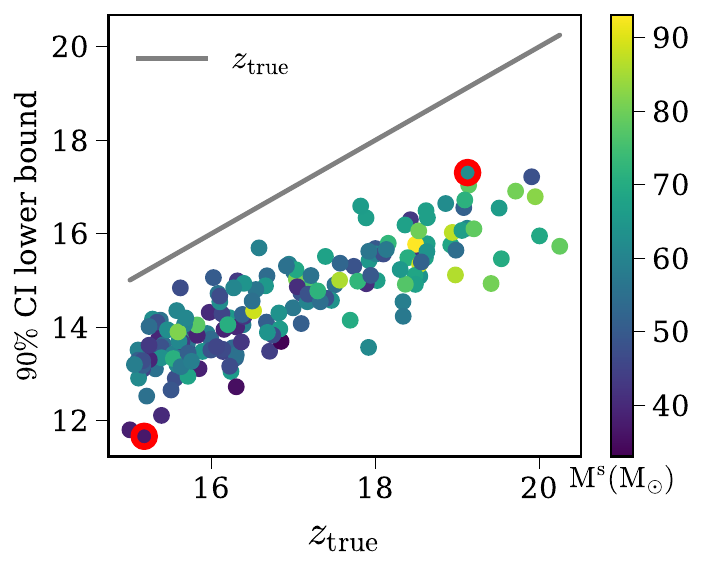}
    \caption{The true source redshift, $\mathrm{z_{true}}$, is plotted against the lower bound of the $90\%$ credible interval from our full Bayesian inference. The color bar encodes the source-frame total mass. Black dots indicate the sources with the highest and lowest lower-bound values, corresponding to the most and least confidently constrained redshifts in our sample. The diagonal black line is the reference line, meaning, a point near the diagonal means the Bayesian inference can confidently place the source close to its true redshift, while points below the diagonal indicate the measurement is less constraining. Points above the dashed line indicate that the lower bound on the infered redshift value is greater than the true value.  The  left plot assumes $f_{\mathrm{low}}=5\mathrm{Hz}$ and the right plot $f_{\mathrm{low}}=10\mathrm{Hz}$.}
    \label{fig:pos_5Hz_90CI_LowerBound_redshift}
\end{figure*}

\section{Results}\label{sec:results}
\label{sec:results}
Performing full Bayesian inference on the detected binaries of our Pop III remnant population described in Sec.~\ref{sec:pe}, we discuss the main results focusing on redshift, component masses, in-plane and out-of plane spins and sky localization uncertainties. 

\subsection{Redshift inference horizon}
\label{sec:redshift}
Establishing the presence of a high-redshift BBH population requires robust constraints on source redshifts. 
Following Ref.~\cite{Mancarella:2023ehn}, for each detected binary we define the one-sided lower credible bound $z_{c\%}$,
\begin{equation}
    P\bigl(z \geq z_{c\%} \mid d\bigr) = \int_{z_{c\%}}^{\infty} p\bigl(z \mid d\bigr)\,\mathrm{d}z = \frac{c}{100},
\label{eq:zlow}
\end{equation}
where $p(z \mid d)$ is the marginal posterior for the source redshift. 
The source then lies at $z \geq z_{c\%}$ with $c\%$ posterior probability. 
Unless stated otherwise, we use $c=90$.

The inferred lower bound of the 90\% CI of the redshift as a function of the true redshift $z_{\rm true}$ is shown in Fig.~\ref{fig:pos_5Hz_90CI_LowerBound_redshift}. 
The left and right panels show the results for $f_{\rm{low}}=\rm{5\rm{Hz}}$ and $f_{\rm{low}}=\rm{10\rm{Hz}}$, respectively. 
The color corresponds to the injected source-frame total mass $M^s$ in solar masses. 

Overall, the inferred lower bounds broadly track the injected redshifts. 
The 90\% CI lower bound for most sources lies below the diagonal, as expected for a one-sided credible bound, but remains close enough to it indicating that a large fraction of the population can be confidently placed at high redshift. 
We find no clear dependence on the source-frame total mass. 
The two frequency cutoffs yield qualitatively similar results, though the bounds obtained for $f_{\rm low} = 5\,\rm Hz$ generally lie closer to the injected values, consistent with the additional low-frequency content improving the redshift constraints.

In Fig.~\ref{fig:pos_5Hz_90CI_LowerBound_redshift}, the red circles mark the binaries with the lowest and highest $z_{90\%}$ estimates. 
For $f_{\rm low}=5\,\mathrm{Hz}$, the lowest bound is $z_{90\%}=12.8$ for a source at $z_{\rm true}=15.1$;
for $f_{\rm low}=10\,\mathrm{Hz}$, it is $z_{90\%}=12.5$ at $z_{\rm true}=15.2$. 
Although the two bounds are comparable, they arise from different binaries: 
in the $5\,\rm Hz$ case from a strongly precessing binary with detector-frame component masses are $526\,M_{\odot}$ and $474\,M_{\odot}$ and a network SNR of $32.5$;
in the $10\,\rm Hz$ case from a binary with little precession, detector-frame component masses of $473.5M_{\odot}$ and $525.5M_{\odot}$, and a network SNR of $32.4$.

These minimum values are themselves informative.
Across the selected catalogue, the smallest $90\%$ lower redshift bound is $z_{90\%}=12.8$ ($12.5$) for the $5\,\mathrm{Hz}$ ($10\,\mathrm{Hz}$) configuration.
Every event in our $z\geq15$ sample therefore has at least $90\%$ posterior probability of lying beyond $z\simeq12$, placing this high-SNR subset in the cosmic-dawn regime where the contribution from Pop~I/II star formation is expected to be subdominant~\cite{Santoliquido:2023wzn}.
This redshift information confidently identifies the sources as high-redshift merger candidates, but does not by itself establish a Pop~III origin.
Distinguishing Pop~III remnants from primordial BHs or other high-redshift populations requires a population-level analysis of the merger-rate evolution and the joint mass, spin, and redshift distributions~\cite{Plunkett:2025mjr}.

The strength of these individual-event bounds partly reflects our restrictive network-SNR threshold of $30$. 
Adopting a lower selection threshold would result in sources with substantially broader redshift posteriors, introducing additional confusion. 
Nevertheless, our results demonstrate that an XG network can isolate a high-confidence sample of cosmic-dawn mergers suitable for population-level studies of their origin.

\begin{figure*}[ht!]
    \centering
    \includegraphics[width=\textwidth]{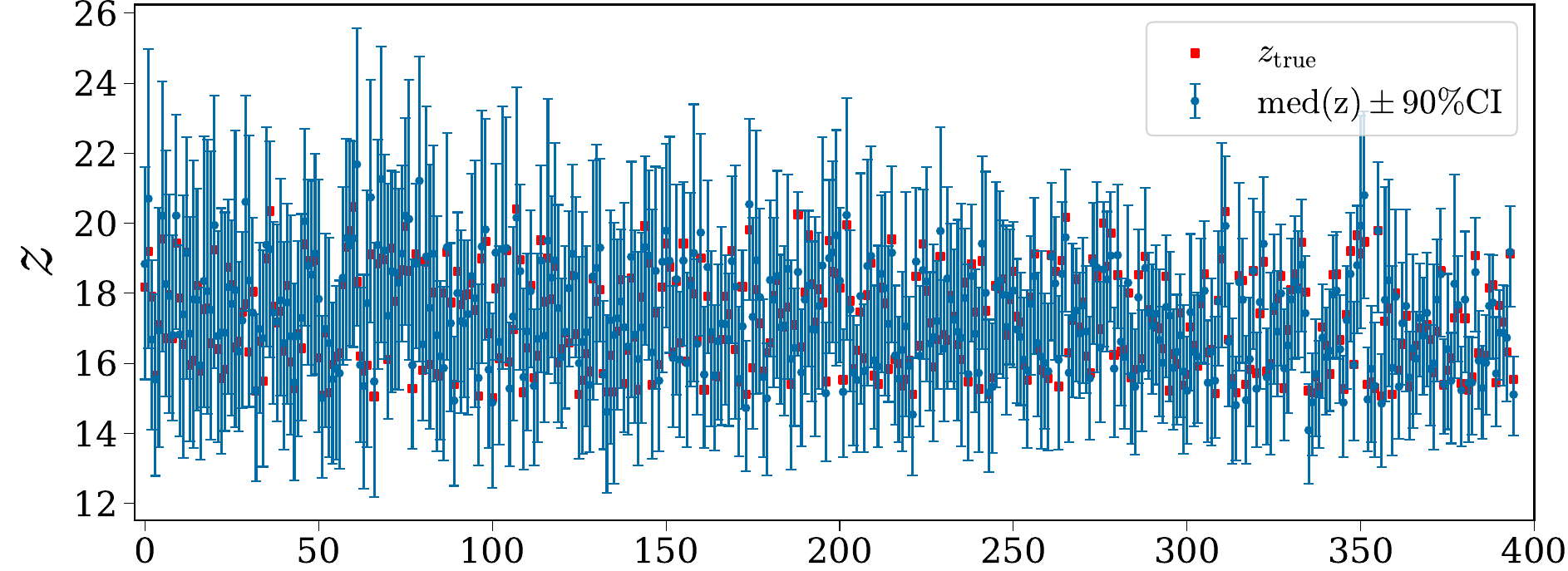}
    \caption{90\% credible bounds on the redshift for all binaries detected by the XG network with a lower cutoff frequency of 5 Hz. The binaries are arranged in increasing order of network SNR. Median values are indicated by blue filled circles. The red rectangles represent the true injection parameters for each binary. The errors bars shrink as the SNR increases, which is visible from the overall envelope.}
    \label{fig:z_errror}
\end{figure*}

\begin{figure*}
     \centering
    \includegraphics[width=\textwidth]{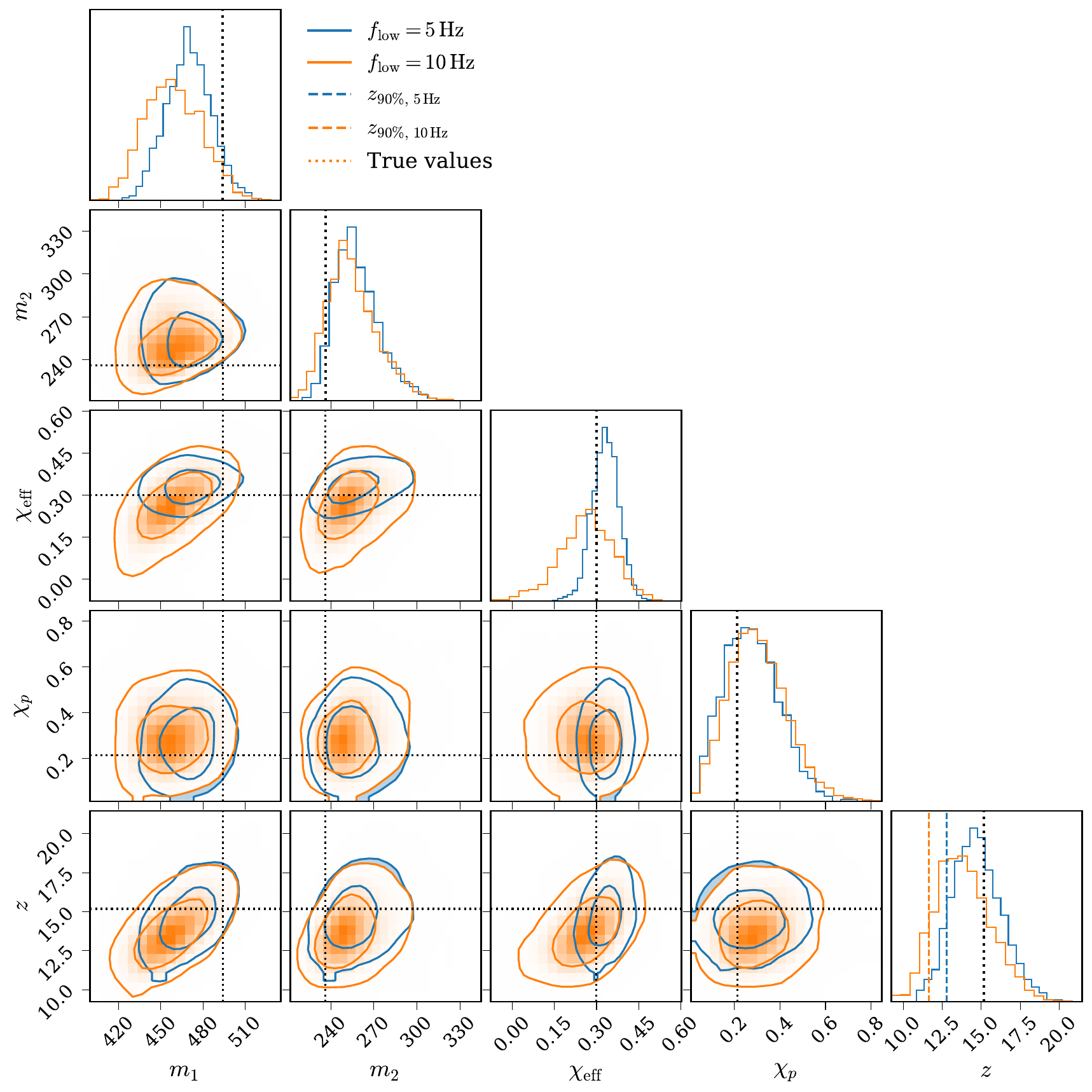}
     \caption{Properties of the system corresponding to the lowest $z_{90\%}$ estimate with $10\, \rm Hz$, where $z_{90\%}=12.5$ at $z_{\rm true}=15.2$, is compared with the same binary analyzed with $5\, \rm Hz$. We show the posterior distributions for the component masses, the effective aligned-spin parameter $\chi_{\rm eff}$, the effective precession spin parameter $\chi_p$, and the redshift for the $f_{\rm low}=5\,\mathrm{Hz}$ (blue) and $f_{\rm low}=10\,\mathrm{Hz}$ (orange) configurations. The dotted lines indicate the true injected parameters. The inferred $z_{90\%}$ values for both configurations are shown as dashed vertical lines in the redshift posterior. }
     \label{fig:corner_z_min_m1_chieff}
 \end{figure*}

Being able to confidently place a binary at a redshift greater than $z \sim 12$ and accurately measure its source properties has potential implications for linking the remnants of the first stars and the formation of BH seeds that later grow into the supermassive BHs observed in high-redshift galaxies. In particular, such measurements can provide access to the population of compact objects that act as seeds or tracers of early BH assembly in the same cosmic epoch that hosts systems such as UHZ1 ($z \sim 10.1$) and GHZ9 ($z \sim 10.4$), the most distant active galactic nuclei observed electromagnetically to date. These systems are often interpreted as being powered by rapidly growing massive BHs, potentially originating from $\sim 10^2$--$10^4\,M_{\odot}$ seed BHs formed from the first generation of stars or direct-collapse channels, and subsequently evolving into $\sim 10^7\,M_{\odot}$ BHs through accretion~\cite{Latif:2025tva, Urrutia:2024hwc}.

Turning now to the highest inferred $z_{90\%}$, the red circle in the upper-right corner of Fig.~\ref{fig:pos_5Hz_90CI_LowerBound_redshift} denotes the most distant source identified in the $f_{\rm low}=5,\mathrm{Hz}$ and $f_{\rm low}=10\,\mathrm{Hz}$ configurations. 
This estimate is obtained from a binary with $z_{\rm true}=19.8$, for which we find a $90\%$ lower bound of $z_{90\%}=18.5$. 
For the $f_{\rm low}=10\,\mathrm{Hz}$ configuration, the largest inferred redshift value is $z_{90\%}=17.5$ for a binary at $z_{\rm true}=19.1$. 
The $f_{\rm low}=5\,\mathrm{Hz}$ configuration yields a systematically larger redshift reach as expected from the increase in the accumulated SNR. 
We note that events with the smallest or largest inferred lower redshift bounds do necessarily maximize (minimize) the absolute difference between the true and inferred redshifts, nor do they guarantee that the posterior peak aligns with the true redshift.
A summary of the 90\% redshift posterior widths, medians and the true redshift for all binaries detected by the $f_{\rm low}=5$ Hz configuration is shown in Fig.~\ref{fig:z_errror}. The binaries are arranged in increasing order of network SNR. 
In particular, for the loudest sources the lower bound lies within $0.1$ of the true value, i.e., $|z_{\rm true}-z_{90\%}|\lesssim0.1$, providing an excellent redshift measurement.

\begin{figure}
    \includegraphics[width= 0.48 \textwidth]{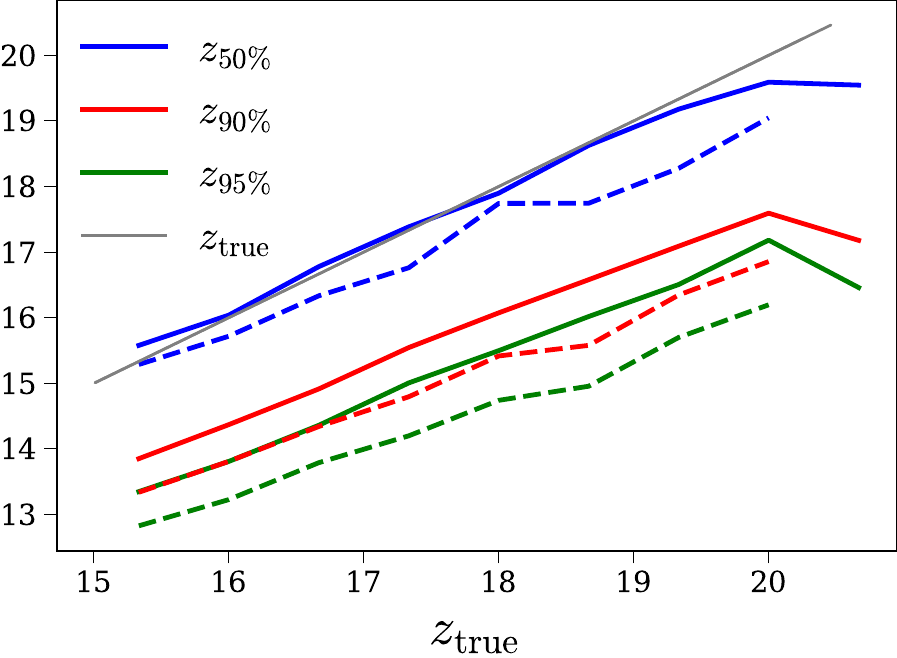}
    \caption{The true source redshift, $\mathrm{z_{true}}$, is plotted against the lower bounds calculated at $95\%$, $90\%$ and $50\%$ credible intervals, considering $f_{\mathrm{low}}=5\mathrm{Hz}$ (solid) and $f_{\mathrm{low}}=10\mathrm{Hz}$ (dashed).  The diagonal line represents the true redshifts.}
    \label{fig:pos_10Hz_CI_LowerBound_redshift_Bands}
\end{figure}

Since $z_{90\%}$ characterises the tail of the posterior, it will necessarily be sensitive to the choice of distance prior. 
Appendix~\ref{ap:resam} compares the lower bounds obtained with the $d_L^{2}$ sampling prior and the lower bounds obtained after reweighting to a uniform merger rate per comoving volume. 
We find that reweighting shifts $z_{90\%}$ downward, corresponding to a less constraining inference of the redshift.  
However, neither of these priors corresponds to the Pop~III merger-rate density itself, which peaks at $z \sim 16$--$18$ for the model adopted here. 
A more self-consistent treatment would be to infer the source redshifts jointly at the population level, which we leave to future work.

The detector-frame component masses, effective spins and redshift posteriors of the binary with the lowest $z_{90\%}$ estimate obtained with the $10\rm Hz$ configuration is shown in Fig.~\ref{fig:corner_z_min_m1_chieff}, and directly compared against the corresponding $f_{\rm low}=5\rm Hz$ results (blue). 
The true injected values are indicated by black dotted lines, while the inferred $z_{90\%}$ values are marked by dashed vertical lines in the marginal redshift posterior. We find that the primary mass, the effective inspiral spin parameter and the redshift are better constrained and more consistent with the injected value when the $5\mathrm{Hz}$ configuration is considered.

In addition to the $90\%$ lower bound discussed so far, we also explore the credible intervals at different confidence levels. We show the $95\%$, $90\%$, and $50\%$ bounds of the redshift for both lower frequency cutoffs in Fig.~\ref{fig:pos_10Hz_CI_LowerBound_redshift_Bands}. As expected, the lower bounds move closer to the diagonal line $z = z_{\rm true}$ as we transition from the $95\%$ to the $50\%$ credible intervals, reflecting the fact that lower-confidence intervals probe progressively higher-posterior-density regions and therefore concentrate more tightly around the most probable values. Similar to the trend reported in ~\cite{Mancarella:2023ehn}, as the true redshift increases the lower bound tends to return lower redshift estimates. 
Moreover, the $f_{\rm low}=10\,\mathrm{Hz}$ estimates (dashed) lie below the $f_{\rm low}=5\,\mathrm{Hz}$ estimates (solid), indicating the importance of having better low-frequency sensitivity for XG detectors to confidently characterise high-redshift binaries. 

\begin{figure}
    \includegraphics[width= 3. in]{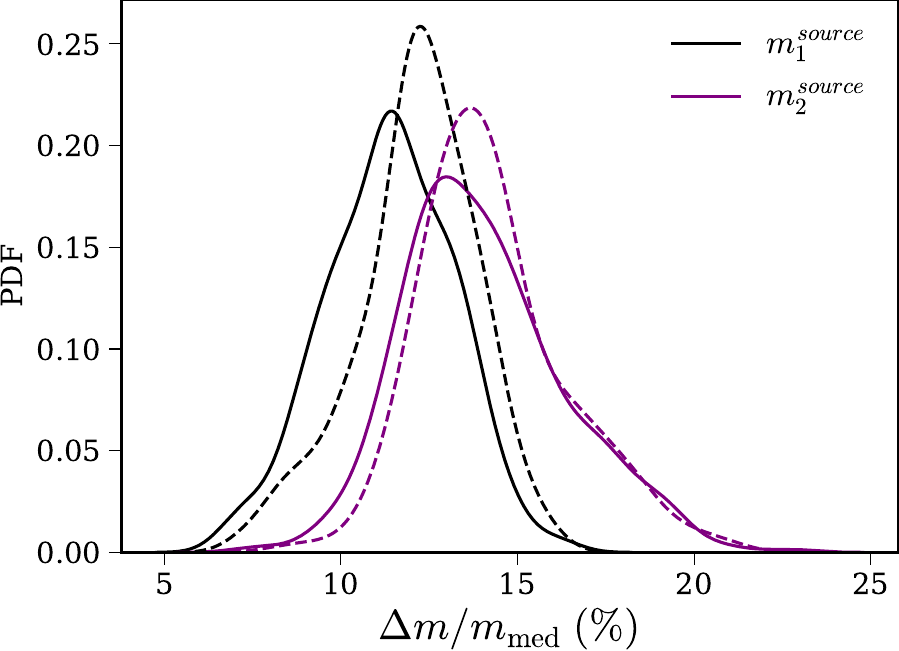}
    \caption{Relative uncertainties (in percent) on the source-frame component masses for the full binary population. Solid and dashed histograms correspond to the $f_{\rm low}=5\mathrm{Hz}$ and $f_{\rm low}=10\mathrm{Hz}$ configurations, respectively.}
\label{fig:m_s_rel_errors}
\end{figure}
\begin{figure*}
    \includegraphics[width=0.48\textwidth]{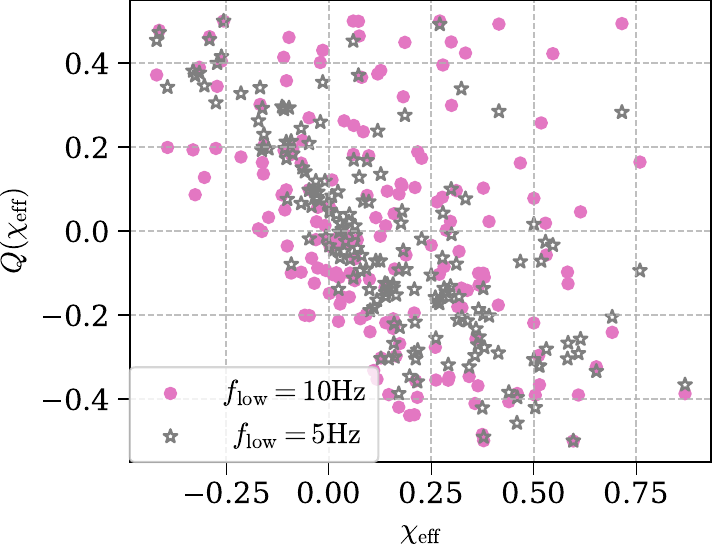}
    \includegraphics[width=0.48\textwidth]{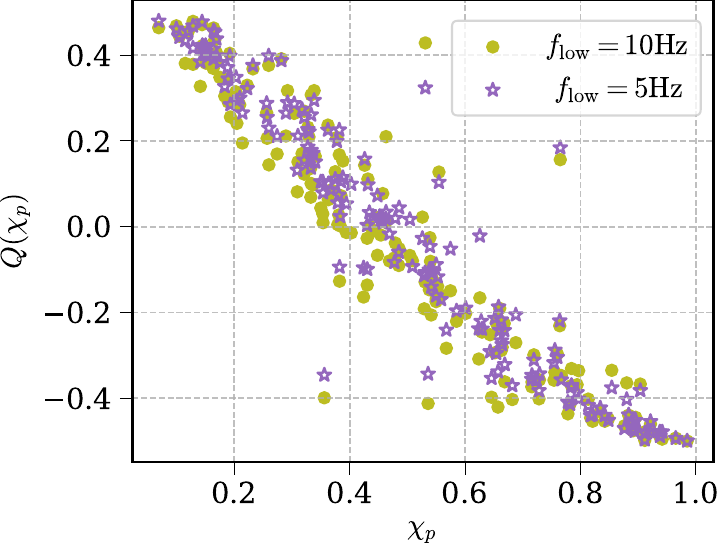}
\caption{
Comparison of the posterior quantile diagnostic $Q$ for the two effective spin parameters, $\chi_{\rm eff}$ (left) and $\chi_p$ (right), as a function of the injected value for the two different low-frequency cutoff for the 175 common binaries detected by both configurations. Values of $Q=0$ indicate unbiased recovery of the posterior median, while positive (negative) values indicate systematic overestimation (underestimation) of the parameter.
}
\label{fig:chi_eff_Q}
\end{figure*}
\begin{figure}
    \includegraphics[width=\columnwidth]{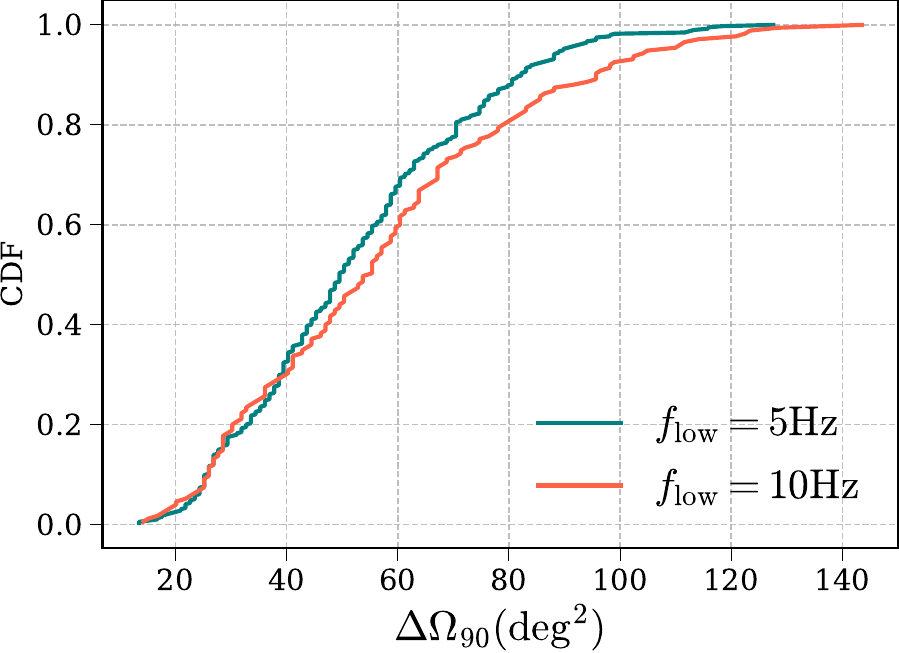}
    \caption{Cumulative distributions of the 90\% sky localization errors in square degrees for the binaries detected with $f_{\mathrm{low}}=5$ Hz (green) and $f_{\mathrm{low}}=10$ Hz (orange). Roughly $60\%$ of the binaries can be localized to within $60\, \rm deg^2$ uncertainty.}
    \label{fig:sky_area_comp}
\end{figure}
\subsection{Uncertainties in source-frame masses}
\label{sec:mass-uncertainty}
Accurate source-frame mass measurements for the most resolvable high-redshift binaries are important for distinguishing between different formation channels and BH seed models, enabling comparisons with population models, and complementing electromagnetic and cosmological observations. In conjunction with redshift uncertainty estimates, it is therefore important to quantify how precisely the source-frame masses of the high-$z$ binaries in our mock population can be measured. Figure~\ref{fig:m_s_rel_errors} shows the relative percentage error (see Eq.~\eqref{eq:relerror}) in source-frame component masses for $f_{\rm low}=5$ Hz (solid) and $f_{\rm low}=10$ Hz (solid). 
The source-frame component masses of the high-redshift binaries can on average be measured to within an uncertainty of $\sim 11\%$ and $\sim 13\%$ with $f_{\rm low}=5$ Hz. 
Compared to the 5 Hz estimates, the 10 Hz uncertainty is slightly larger as expected.
We also find that the primary mass is slightly better constrained than the mass of the secondary. The detector-frame mass measurement is summarised in Appendix~\ref{app:fullresult}. 

This slight improvement can be understood as follows: For $f_{\rm low}=10\,\mathrm{Hz}$, only the final few GW cycles are in band, and the likelihood is dominated by the merger and ringdown, as illustrated in Fig.~\ref{fig:char_freqs}. 
The frequencies of these features depend most directly on the detector-frame remnant mass and spin, while the relative amplitudes and phases of the multipoles retain some sensitivity to the progenitor mass ratio and spins~\cite{Kamaretsos:2012bs,London:2014cma,Borhanian:2019kxt,Zhu:2023fnf,Caceres-Barbosa:2025lhr}. 
With little inspiral evolution in band, different combinations of mass ratio and component spins can produce similar remnant properties and merger-ringdown morphology. 
The mass ratio and spins therefore remain strongly correlated, limiting our ability to accurately recover the individual component parameters.
Extending the analysis to $5\,\mathrm{Hz}$ brings more of the late-inspiral and merger evolution into the sensitivity band of the detector, providing additional information with which to distinguish these configurations. 
The resulting improvement is clearest for the mass and redshift measurements discussed above, but does not translate into a significant improvement in the spin measurement (see below). 

\subsection{Relative uncertainties in spin parameters}
\label{sec:spin-uncertainty}

Accurate inference of spin parameters is essential for understanding the formation and evolution pathways of compact binaries. To assess systematic biases in their recovery, we use the posterior quantile diagnostic $Q$ defined in Eq.~\eqref{eq:Qdef}, which quantifies the offset between the true parameter value and the posterior distribution. 
In particular, we consider the effective inspiral spin parameter $\chi_{\rm eff}$~\cite{Ajith:2009bn} and the precession spin parameter $\chi_{p}$~\cite{Schmidt:2014iyl}, which encode the aligned-spin contribution to the spin dynamics and the in-plane spin components responsible for orbital precession, respectively. 
Figure~\ref{fig:chi_eff_Q} shows $Q(\chi_{\rm eff})$ (left) and $Q(\chi_p)$ (right) as functions of the injected values for the $f_{\rm low}=5\,\mathrm{Hz}$ (unfilled stars) and $f_{\rm low}=10\,\mathrm{Hz}$ (filled circles) configurations.
Deviations of $Q$ from zero indicate systematic offsets between the injected value and the posterior median, while the distribution of $Q$ across the population tells us about the calibration of the posteriors. 
For well-calibrated inference, we would expect that $Q$ is uniformly distributed on the interval $[-1/2,1/2]$~\cite{Cook:2006cal,Talts:2018}.

For $\chi_{\rm eff}$, the $5\,\mathrm{Hz}$ analysis yields $Q$-values somewhat more concentrated around zero than the $10\,\mathrm{Hz}$ analysis, indicating that the posterior medians lie closer to the injected values, though this does not by itself imply smaller statistical uncertainties. The reduced spread, however, suggests that a lower cutoff frequency of 5 Hz leads to a slightly more informative effective spin measurement. 

For $\chi_p$, both configurations exhibit a pronounced anti-correlation between $Q$ and the injected value. 
Small values of $\chi_p$ are systematically overestimated and large $\chi_p$ underestimated, consistent with posteriors that are pulled toward the bulk of the prior when the data are only weakly informative about precession.
Lowering the cutoff to $5\,\mathrm{Hz}$ does not remove this trend.

Astrophysically, this implies that we will only be able to obtain limited spin information about the high-redshift compact binary population. 
This is an important limitation for distinguishing formation channels, whose mass distributions can be highly degenerate. 
For example, Ref.~\cite{Costa:2023xsz} finds similar mass distributions for Pop I/II stars and Pop III remnants. 
However, the two populations should differ in spin, as Pop III stars are thought to form via rapid disk accretion and, lacking the metal-line-driven winds that brake rotation, to be born with high natal spins~\cite{Yoon:2012pop,Stacy:2013pop1,Stacy:2013pop}. 
Our results suggest that this discriminating power will remain limited for high-redshift sources, even with a lower cutoff frequency, so that population-level constraints on the origin of these binaries will need to rest primarily on masses and merger rates.

%
\subsection{Sky localization uncertainties}
\label{sec:skyloc}
Finally, precise sky localization of the most resolvable high-redshift binaries is important both for assessing the feasibility of host-galaxy identification and for enabling cross-correlation with surveys of large-scale structure. 
Of particular interest are cross-correlations with $21$\,cm intensity maps of the cosmic-dawn era, such as those anticipated from SKA~\cite{SKA:2026,Baker:2026rpa}, and joint studies of the origin and evolution of the first galaxies, combining deep galaxy surveys, such as JWST~\cite{JWST:2023NatAst,Naidu:2025xfo}, with GW observations of massive black hole binaries by LISA~\cite{LISA:2024hlh}. 
In this context, measurements of the masses, spins, and merger rate of Pop~III remnants would provide an independent constraint on the light-seed channel of early massive black hole growth, complementing electromagnetic censuses of high-redshift galaxies and AGN and LISA observations of their subsequent mergers.

From the posterior distributions of the right ascension and declination, we construct the 2D sky localization uncertainty $\Delta \Omega_{90\%}$. The cumulative distribution function (CDF) of the 90\% localization errors are shown in Fig.~\ref{fig:sky_area_comp}. 
The details of the calculation are given in App.~\ref{app:sky}.

The difference in the lower cutoff frequency has a negligible impact on the sky localisation of high-redshift sources. There is a slight overall improvement when the cutoff frequency is lowered to $5\,\mathrm{Hz}$. 
Accurate source localization is important for several astrophysical applications. Together with luminosity distance measurements, it reduces the three-dimensional localization volume, enabling statistical host-galaxy identification and cross-correlation with galaxy catalogues for cosmological studies. Furthermore, improved sky localization enables targeted electromagnetic follow-up observations in scenarios where electromagnetic counterparts are expected, maximizing the scientific return of potential multi-messenger campaigns~\cite{MPSAC,Saini:2022hrs}.

%
%
%
\section{Discussion}
\label{sec:conclusions}
In this work, we have quantified the ability of a CE+ET GW detector network to characterise massive BBH mergers at $z\geq 15$.
We consider an astrophysically motivated Pop~III remnant population and perform full Bayesian inference on all binaries that exceed a network SNR of $30$.
For all our analyses we consider two low-frequency cutoffs of the detectors, $f_{\rm low}=5$ Hz and $10\,\mathrm{Hz}$, which represent an optimistic and conservative estimate of the enhanced low-frequency performance of XG detectors. 
Access to the $5$--$10\,\mathrm{Hz}$ band increases the number of detected sources passing by a factor of $\sim2.3$.
It also tightens the redshift constraints, with every event in the selected sample being confidently placed beyond a redshift $z\simeq12$ as characterised by the lower bound of the marginal 90\% CI. 
In particular, for the loudest sources the $90\%$ lower bound lies within $|z_{\rm true}-z_{90\%}|\lesssim0.1$ of the injected redshift.

The source-frame component masses are on average recovered with relative uncertainties of around $11-13\%$ across the population, with the primary mass slightly better constrained than the secondary. The mass measurement is slightly better when considering $f_{\rm low}=5$ Hz, which can be attributed to observing more GW cycles from the late inspiral in band.
The spin measurements, on the other hand, benefit more modestly, remaining limited by correlations between the mass ratio and the component spins that cannot be broken even for $f_{\rm low}=5$ Hz. In particular, we find that the measurement of precession in the considered high-$z$ population is either uninformative or biased. Lowering the cutoff frequency does not remove this trend. As the mass distributions for different high-$z$ BBH populations are degenerate, accurate spin measurements could provide an important pathway to distinguishing between them. However, our results suggest that only limited spin information can be obtained providing limited discriminating power. 

Across our Pop~III remnant population, the 90\% sky-localisation regions required for cross-correlation with high-redshift surveys are below $\sim 140\, \rm deg^2$, but improve only marginally between the two cutoff frequencies.

Mergers at $z=15$--$20$ occur when the Universe is only $\sim0.18$--$0.27\,\mathrm{Gyr}$ old, leaving little time for binary formation and evolution before merger.
A catalogue of such systems would probe the prompt end of the delay-time distribution and constrain the fastest pathways through which the first stars produced merging BHs~\cite{Liu:2024mkh}.
Our $90\%$ lower bounds place every selected source beyond $z\simeq12$, in a regime where Pop~I/II star formation is expected to be subdominant~\cite{Santoliquido:2023wzn}, providing important constraining power on early-universe BH mergers. 
Although redshift alone cannot distinguish a Pop~III remnant from a primordial BH, the improved distance measurements from the $5$--$10\,\mathrm{Hz}$ band will strengthen population-level constraints from the merger-rate evolution and the joint mass, spin, and redshift distributions.

We note that our redshift uncertainties are purely statistical, and we have assumed that the observed signals are unlensed.
At such cosmological distances, weak lensing can introduce a scatter on the inferred luminosity distance that grows with redshift, reaching $\mathcal{O}(10\%)$ at $z \gtrsim 15$~\cite{Hirata:2010ba, Dai:2016igl}, which is comparable to the statistical distance uncertainties of the loudest sources in our sample.
In addition, strongly magnified sources at lower redshift can scatter into the apparent high-redshift tail of the observed population, contaminating any candidate sample of high-redshift mergers~\cite{Dai:2016igl,Oguri:2018muv}.
The redshift bounds quoted here should therefore be treated as statistical estimates, as they do not account for lensing.

The Bayesian framework used here, based on the \texttt{IMRPhenomXPHM} waveform model, carries over directly to other proposed early-Universe populations, including primordial BHs.
Our results also make it clear that characterising compact binary populations in the early Universe with GWs can benefit from improved sensitivity in the $5$--$10\,\mathrm{Hz}$ band. 
Sensitivity over this range will therefore be important for measuring the redshift evolution and intrinsic properties of the earliest BH merger populations in the Universe.

\section*{Acknowledgments}
The authors would like to thank Alberto Vecchio for useful discussions.
N.V.K., G.P and P.S. acknowledge support from STFC grant ST/Y00423X/1;
G.P. and P.S. also acknowledge support from STFC grants ST/V005677/1 and UKRI2493.
G.P. is very grateful for support from a Royal Society University Research Fellowship URF{\textbackslash}R1{\textbackslash}221500 and RF{\textbackslash}ERE{\textbackslash}221015. GP gratefully acknowledges support from an NVIDIA Academic Hardware Grant. P.S. also acknowledges support from a Royal Society Research Grant RG{\textbackslash}R1{\textbackslash}241327. 


\newpage

\begin{widetext}

\appendix
\section{Cosmological Reweighting and Resampling}
\label{ap:resam}
We resample the binary population to account for cosmological volume effects using a $\Lambda$CDM cosmology consistent with Planck measurements, with parameters $H_0 = 67.9\,\mathrm{km\,s^{-1}\,Mpc^{-1}}$, $\Omega_m = 0.3065$, and $\Omega_\Lambda = 0.6935$. For each sample, the redshift $z$ is obtained by inverting the luminosity distance $D_L$ using the cosmological relation $D_L(z)$, using \texttt{Astropy}~\cite{astropy:2013}. 

We construct an importance-sampling weight that reweights posterior samples, originally drawn under a prior uniform in luminosity distance, into a redshift distribution consistent with a $\Lambda$CDM cosmology and an optional star-formation-rate (SFR) evolution of the form $\psi(z)\propto(1+z)^{\kappa}$.
The weight is proportional to the ratio of
the target redshift prior to the sampling prior, and takes the form
\begin{equation*}
w(z, d_L) \propto
\frac{1}{(1+z)^{2-\kappa}\left[E(z)\,\dfrac{D_L}{D_H} + (1+z)^{2}\right]},
\end{equation*}
where $E(z) = H(z)/H_0$, $D_H = c/H_0$ is the Hubble distance, and the bracketed term arises from the Jacobian $\mathrm{d} D_L / \mathrm{d}z$. 
Here, $\kappa$ controls the redshift dependence of the star formation rate evolution, such that setting $\kappa = 0$ recovers a merger rate uniform in comoving volume and source-frame time. 
The weights are normalized and used to resample the posterior.

Figure~\ref{fig:dist_z_resampling_comparison_pos_0} shows the posterior distributions of luminosity distance and redshift for a representative binary, before and after cosmological reweighting.
Figure~\ref{fig:pos_5Hz_10Hz_rempling} compares the $90\%$ lower bounds on redshift before (red) and after (blue) resampling, with the reweighting shifting the lower bounds to slightly smaller values. 
The overall effect of the reweighting is modest for our fiducial merger-rate evolution, suggesting that our results are largely robust to the choice of distance prior.

\begin{figure*}
    \includegraphics[width= 5. in]{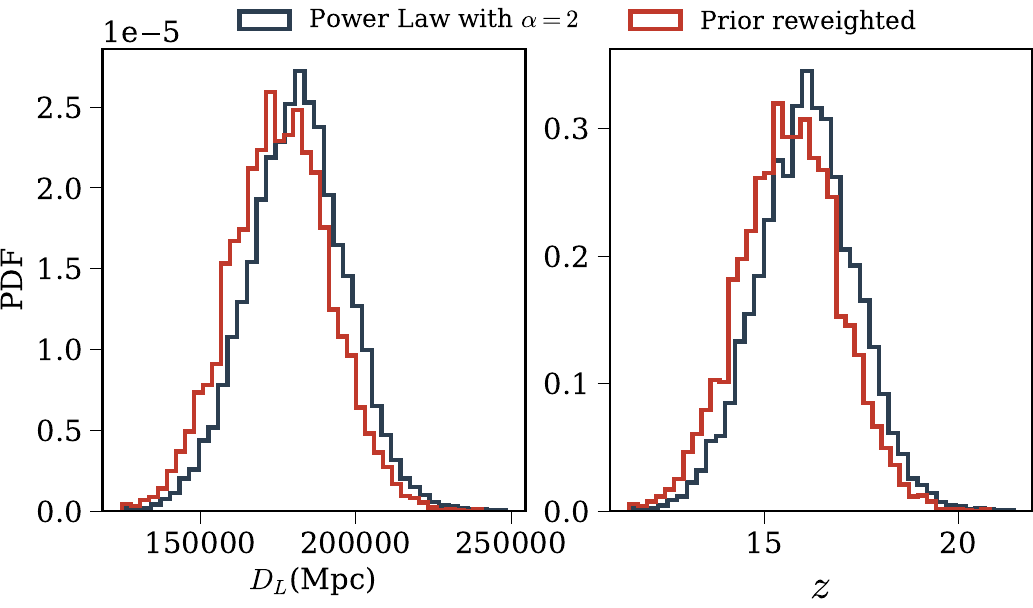}
    \caption{The posterior distributions on luminosity distance and redshift from a representative binary system, before and after resampling to account for the cosmological reweighting.}
    \label{fig:dist_z_resampling_comparison_pos_0}
\end{figure*}
\section{Sky localization error estimates}
\label{app:sky}
The sky localization uncertainty is quantified by constructing a credible region on the celestial sphere using HEALPix pixelization~\cite{Gorski:2004by}. Posterior samples in right ascension ($\alpha$) and declination ($\delta$) are mapped to spherical coordinates via
\begin{equation}
\theta = \frac{\pi}{2} - \delta, \qquad \phi = \alpha,
\end{equation}
and assigned to HEALPix pixels at a chosen resolution $N_{\rm side}$~\cite{Gorski:2004by}. A discrete probability distribution over pixels is then constructed from the sample counts. The pixels are ranked in descending order of probability, and the smallest set enclosing a fraction $c$ of the total probability is selected. The corresponding sky area is computed as
\begin{equation}
\Delta \Omega_c = N_{\rm pix} \times \Omega_{\rm pix},
\end{equation}
where $N_{\rm pix}$ is the number of selected pixels and $\Omega_{\rm pix} = 4\pi / N_{\rm pix}^{\rm tot}$ is the solid angle per pixel, with $N_{\rm pix}^{\rm tot} = 12\,N_{\rm side}^2$. This defines the $c\%$ credible sky localization region~\cite{Singer:2015ema}.
\begin{figure*}
    \includegraphics[width= 5. in]{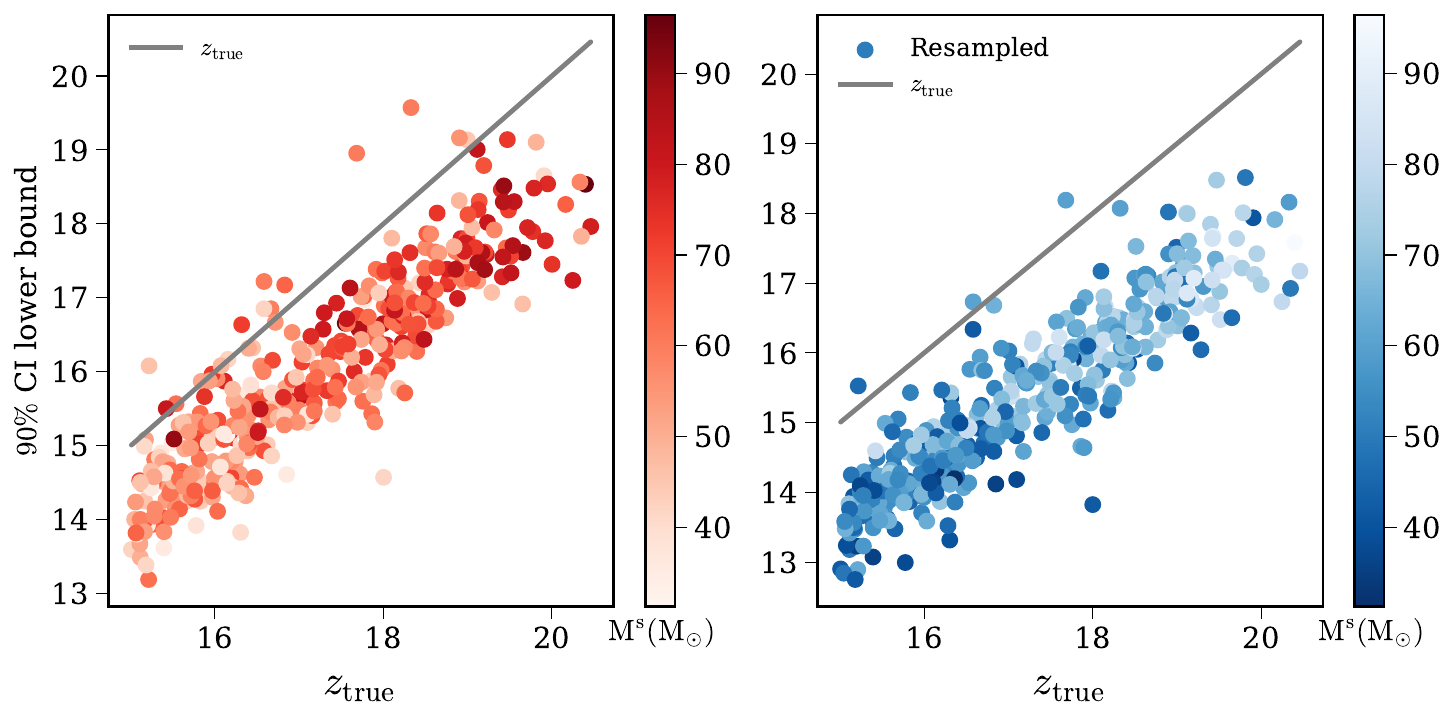}
    \includegraphics[width= 5. in]{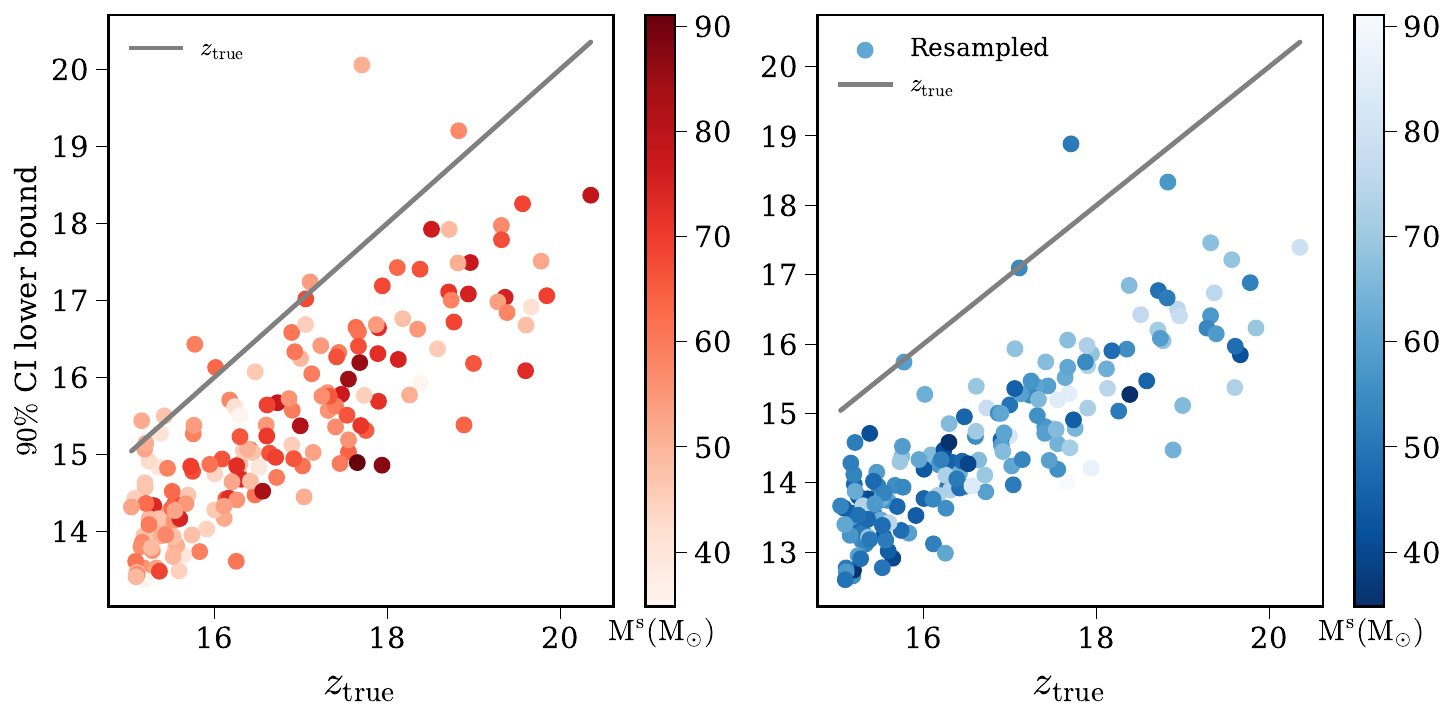}
    \caption{Comparison of the $90\%$ lower bound redshift estimates before and after resampling the redshift posteriors. The red and blue are estimated from raw and resampled redshift posteriors respectively. $\mathrm{f_{low}}$ 5 and 10 $\rm Hz$ are shown in the top and bottom panels.}
    \label{fig:pos_5Hz_10Hz_rempling}
\end{figure*}
\section{Error bars on redshift and component masses from the full population}
\label{app:fullresult}
Figure~\ref{fig:pos_5Hz_error_bar_plot} shows the $90\%$ bounds on the redshift (first row), detector frame primary mass (second raw) and detector frame secondary mass (third row). The binaries are arranged in the increasing order of network SNR, where the sources on the left are weaker compared to the loud sources in the right. Median values are indicated by blue filled circle. The red rectangles represent the true injection parameters for each case. The errors bars on redshift reduces as the SNR increases, which is visible from the overall envelope of the plot.
\begin{figure*}
    \includegraphics[width= 5.5 in]{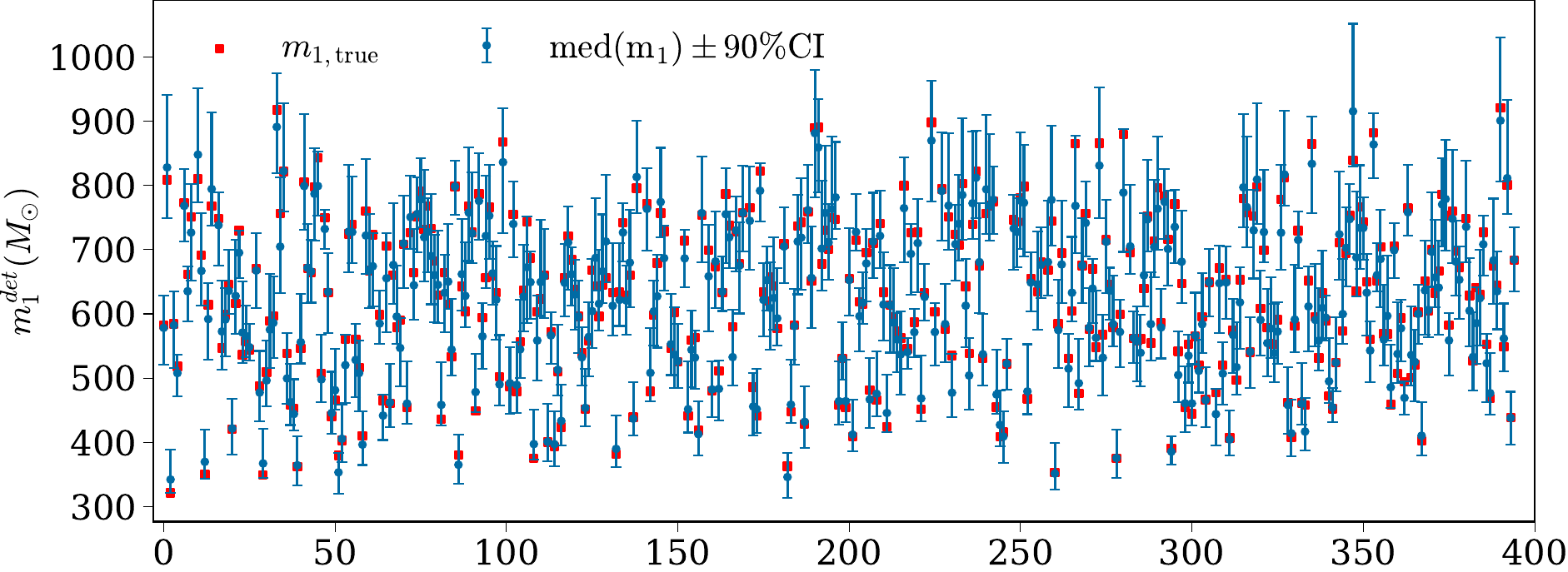}
    \includegraphics[width= 5.5 in]{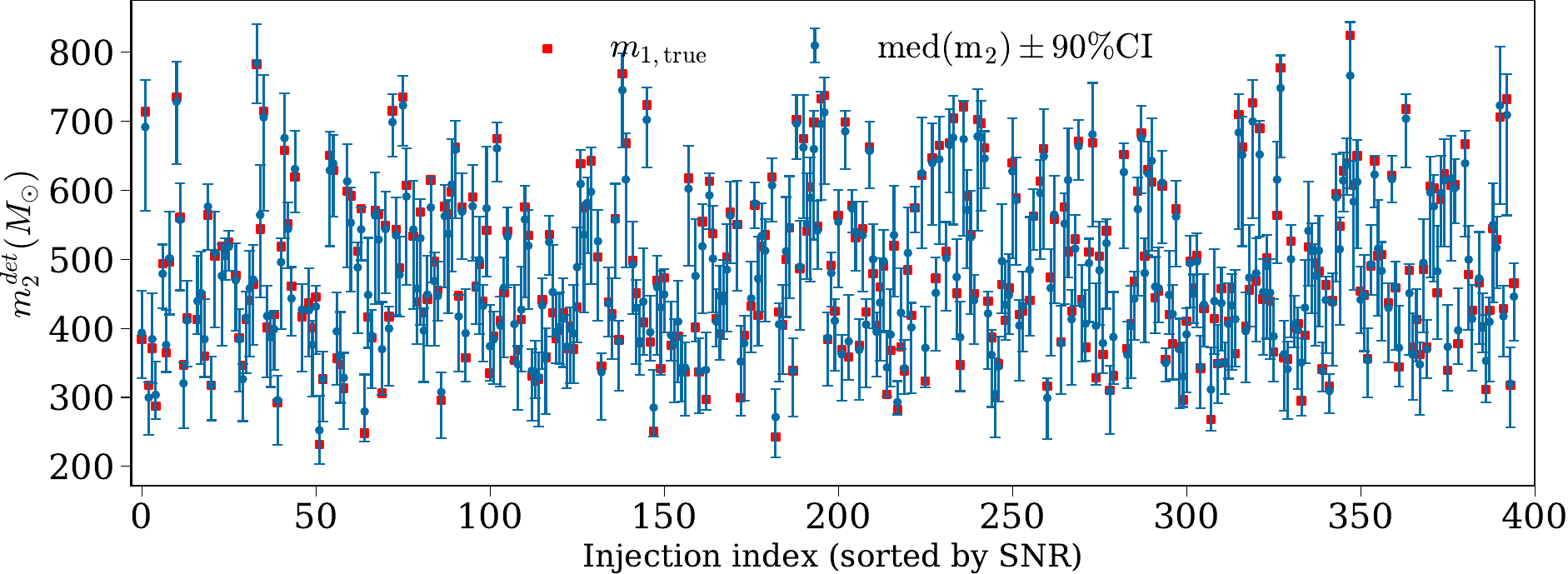}
    \caption{Error bars on the detector-frame primary and secondary mass for the analysis with $f_{\rm low} = 5$ Hz. The sources are arranged in increasing order of their network SNR in the XG detector network. The red squares mark the injected value, the blue circle the posterior median with the error bars indicating the 90\% CI.}
    \label{fig:pos_5Hz_error_bar_plot}
\end{figure*}


\end{widetext}

\clearpage
\bibliography{Refs}
\end{document}